# PCM-net: A refractive index database of chalcogenide phase change materials for tunable nanophotonic device modelling

**Hyun Jung Kim[1,2*], Jung-woo Sohn[3], Nina Hong[4], Calum Williams[5], William Humphreys[2]**
1. National Institute of Aerospace, Hampton, Virginia 23666, USA
2. NASA Langley Research Center, Hampton, Virginia 23666, USA
3. Institute for Life Science Entrepreneurship, New Jersey 07083, USA
4. J.A. Woollam Co., Inc, Lincoln, Nebraska 68508, USA
5. Department of Physics, Cavendish Laboratory, University of Cambridge, Cambridge, CB3 0HE, UK
* corresponding author: hyunjung.kim@nasa.gov

## Abstract

The growing demand for multifunctional nanophotonic devices has led to the exploration, and utilization, of a plethora of exotic electro-optical materials. Recently, chalcogenide glass based phase change materials (PCMs) have shown utility as a tuning material for a range of nanophotonic devices. Owing to their low loss, ultrafast switching speeds and wide waveband operation, PCMs are integrated in an increasing number of next-generation tunable components, including integrated photonic switches, metasurface optics and tunable spectral filters. Nonetheless, modelling of PCM-based devices is challenging—both in terms of accurate representation of experimentally-derived material properties in different phase states, and standardization of results across the research community. Further, as each device requires optimization of specific performance metrics dependent on their respective application, any inaccuracies will lead to erroneous outcomes. In this work, we introduce PCM-net (http://nekocloud.com/pnet/): an online database of the complex refractive indices of a variety of chalcogenide glass PCMs (such as GeSbTe), as an accessible and indexed repository for data sharing across the PCM community. Refractive indices ($n$) and extinction coefficients ($k$) between amorphous and crystalline states are directly extracted from experimentally-derived data in numerous academic research articles, and collated into the material resource database. Due to the inaccuracies associated with our data collection methods, this data is supplemented with additional computationally-generated data, obtained through WVASE®—a commercial ellipsometry analysis software package. To demonstrate the utility of PCM-net, we provide a NASA application-driven device optimization example using the optical properties of PCMs collected with our database. We anticipate the database providing great use to the PCM community and coordinated research efforts enabled by PCM-net will promote the shared repository for the selection of appropriate PCMs for tunable nanophotonic device design for a range of applications.

# Introduction

In recent decades, photonics has matured to a key enabling technology, underpinning a plethora of modern technologies from communications, energy to healthcare [1-3]. A particularly exciting research direction within photonics is tunable and multifunctional solid-state devices—whose optical response can be dynamically controlled post-fabrication. Tunable optical properties, such as phase, wavelength and amplitude modulation are attractive across numerous applications, from optical storage media, reconfigurable integrated photonic elements to spectrally adjustable bandpass filters [4-8]. Underpinning such devices are active optical materials, those which exhibit a change in refractive index through external stimuli [9-11].

With growing demand across different application areas for faster modulation speeds, wider waveband coverage and more cost-effective systems, researchers are exploring a host of 'more exotic' active materials [7-8,12-13]. Further, as devices continually shrink, nanophotonic and metasurface based design methodologies are becoming commonplace [8,12]. As a result, phase change materials (PCMs), and in particular chalcogenide-glass based PCMs such as GeSbTe, have seen renewed research interest in recent years [7-8,14-15]. Owing to their success integrated into low cost rewritable optical storage media, PCMs are now emerging as a potential material of choice for a wider host of next-generation photonic systems. This is due to their ultrafast non-volatile phase transition (amorphous-to-crystalline) resulting in significant low loss refractive index contrast modulation. PCM-based tunable optical devices have been shown capable of providing optical contrast across multiple wavebands, ultrafast (GHz) modulation speeds, and through low energy stimuli [12,16], having wide ranging applications from metasurface optical components to display technology. Irrespective of application, designing reconfigurable optical devices based on PCMs presents some common overarching challenges, for example accurate representation of the complex refractive index and device output optimization across multiple PCM-states. PCMs exhibit strikingly different optical properties (refractive index contrast) when switching from crystalline to amorphous state. With the fact that these properties are highly dependent on both deposition and characterization methodology, even for the same PCM composition, means the device design is strongly susceptible to which model or experimentally-derived set of values are used.

Here, we introduce PCM-net (http://nekocloud.com/pnet/): a convenient online database (library) of different PCMs' optical characteristics (phase-dependent complex refractive indices) obtained through data extraction from academic research articles and computationally generated through commercial WVASE® ellipsometry analysis software. GeSbTe (GST), the widely utilized transition metal chalcogenide alloy, was the main focus for our initial PCM-net implementation. We also present a future outlook to expedite multifunctional application-driven optimization of PCM-based optical devices by supplying extensive PCM resources. Establishing a common PCM optical database is important for researchers designing, simulating and optimizing active PCM-based devices such as metasurfaces for multi-state tunability. Moreover, material characteristics data stored on our PCM database provides useful resources for application-driven and machine learning-based metasurface / nanophotonic designs [17,18]. The inspiration of PCM-net has arisen from the material sciences community, where rapid progress has been made largely mediated through a culture of sharing, benchmarking, and use of common datasets, such as those featured in CIFAR, NASA images, and MetaNet [19-22]. We expect, and hope, PCM-net will promote coordination and collaboration within the materials and photonics community, and that PCM-net can become a repository for collections of PCM materials data utilized for a wide range of device designs.

## Methodology: Data collection of PCM optical characteristics

Refractive index and extinction coefficient values depend on whether the PCM (e.g. GeSbTe) is in its amorphous, crystalline or partial crystallinity state. For a community database, the data can be, (a) collected from experiments directly, (b) extracted from data (graph plots, supplementary datasets etc.) in academic research articles, (c) submitted by research contributors (crowdsourcing), or (d) computationally generated (simulated). Table 1 shows the summarized pros and cons of each data collection method, and we further overview and discuss each approach, including major considerations.

**Table 1**. Pros and cons of data collection methods for material characteristics

| Approach | Pros | Cons | Potential solutions |
| --- | --- | --- | --- |
| *From experiments* | Original source of data  Control on generation of data types | Might need processing and error corrections  Highly dependent on experimental apparatus and methods | Development of data quality checking methods needed |
| *Data extraction from papers* | Availability of quality data  Software tools available for extracting data from chart plots | Time-consuming manual work and data cleaning still needed for the tools | Potential for paper repositories to provide chart plot datasets (metadata) bundled with the paper document file |
| *Crowdsourcing* | Potential for obtaining a large number of contributed datasets | Data submission system needs implementation  Validation or quality assurance of submitted data are required | Development of data quality checking methods needed  Making data submission template in spreadsheet format can serve as a starting step for the development of the data submission system |
| *Simulation* | Capable of generating diverse datasets with control on the data types | Simulated data might show different characteristics than the real-world data depending on the simulation configurations | |

**From experiments**
Experiments are naturally the intuitive method to generate data and spectroscopic ellipsometry (SE) is a well-established technique used to study the dispersive refractive index ($n$) and extinction coefficient ($k$) optical constants of thin-film materials [23-25]. Ellipsometry measures a change in polarization as light reflects or transmits from a material structure. The polarization change is represented as an amplitude ratio ($\Psi$) and the phase difference ($\Delta$), the raw measurement from an ellipsometry. The measured $\Psi$ and $\Delta$ values are analyzed to determine material properties of interest including the optical constants. Since SE is based on the ratio of two measured values, it is accurate and reproducible. Commercially available ellipsometry software such as 'WVASE®' can be modelled using predetermined equations based on the opaqueness, surface roughness, number of layers, etc. of the sample and it can easily fit to the data to determine the results. The model implements a mean square error (MSE) that measures how accurate the model is with respect to fitting the data. Therefore, a model with lower MSE values for fitting the data is considered to be more accurate. Since the optical properties can be represented as the complex dielectric function, the experimentalist needs to control the types of data to be generated with the choices of dielectric functions '$\varepsilon_1(\lambda)$, $\varepsilon_2(\lambda)$', optical constants '$n(\lambda)$, $k(\lambda)$', and so on.

For the PCM-net database, both in-situ and ex-situ ellipsometry measurements were performed. The in-situ measurements were taken while the PCM GeSbTe is placed in a customized vacuum chamber at elevated temperature. These measurements are, in detail:

1) In situ real-time spectroscopic ellipsometry measurements by collaborator Professor Stefan Zollner (and research group) at New Mexico State University [26]. The complex refractive indices and extinction coefficients of Carbon-doped $Ge_2Sb_2Te_5$ (C:GST) layers (750 nm) that are deposited on single-side polished Si wafers with 400 nm $SiO_2$ capping layer were characterized using Variable Angle Spectroscopic Ellipsometry (VASE) over a broad spectral (190 nm – 40 µm). For the broad wavelength test, two different instruments, J.A. Woollam VASE (190 nm – 2.5 µm) and Fourier-Transform Infrared (FTIR) Variable Angle of incidence Spectroscopic Ellipsometer (FTIR-VASE for 1.5 µm – 40 µm) are applied. Utilizing temperature-dependent spectroscopic ellipsometry, the research group investigates the optical constants of C:GST before and after crystallization occurs. The temperature range is set from 80K (-193°C) to 750K (477°C) and the incidence angle is set to 70° in an ultra-high vacuum cryostat. For the data collected from the two different instruments, custom window correction was applied to avoid merging discontinuities. The ellipsometric angles $\psi$ and $\Delta$ were fitted using various analytical models (Tauc-Lorentz, Cody-Lorentz, and Gaussian oscillators) and interpolation schemes (point-by-point or B-spline) to determine the optical constants as a function of wavelength and temperature of the combined dataset.

2) Ex-situ spectroscopic ellipsometry measurement with post-annealed PCMs [7,8]. The complex refractive indices and extinction coefficients of $Ge_2Sb_2Te_5$ (GST) films (100nm thick) deposited on $CaF_2$ substrate were measured over wavelength range of 1 – 10 µm using the J.A. Woollam ellipsometry system. The GST film was annealed at 145°C (418K) and 172°C (445K) inside the vacuum chamber for crystallization (phase change) prior to ellipsometry characterization after the as-deposited amorphous GST film test. Both the J.A. Woollam RC2® ellipsometry system and IR-VASE with the range from 55° to 75° angles of incidence and 10° step were used to measure the optical properties. The VASE v.6.33 and WVASE v3.908 software was used for the data analysis and the model was constructed with Kramers-Kronig consistent GenOsc layer.

**Data extraction from papers**
Material data can be readily obtained from published research articles. Research articles provide the data in the form of graphical plots (visualizations) or alternatively through supplementary datasets (less widely available). Since the former does not directly provide the numerical dataset, it requires a conversion process from visual graph plot images back to the original dataset. While a range of chart recognition tools are available for converting graphical images to numerical datasets, it should be noted that the conversion process is hard to automate and requires manual work. For example, WebPlotDigitizer [27] can generate the original dataset from a given plot image file by extracting the plot line. WebPlotDigitizer automates the process of detecting the color difference between the line and the background, and extracts the plot line. But users still need to manually specify the details of the plot with a mouse or a pen input device, such as the plot line area, x-y axes, and the scale of the axes. Figure 1 describes how users extract a characteristic dataset for a PCM from a plot image using WebPlotDigitizer.

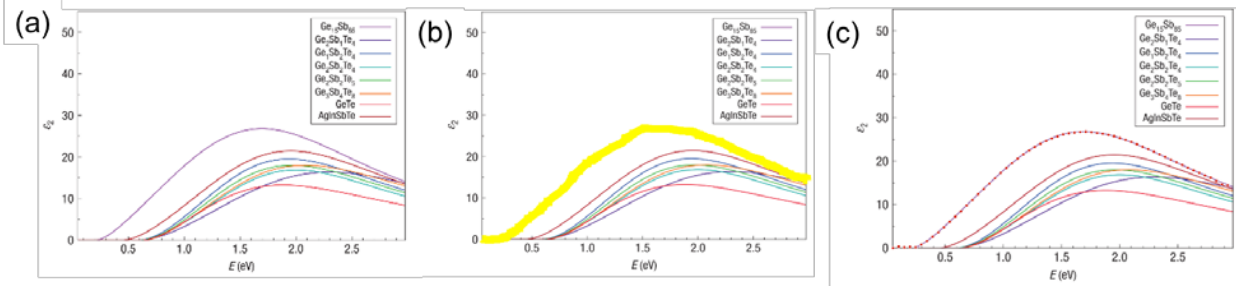

**Figure 1.** Extracting a dataset of constant real part ($\varepsilon_1(\lambda)$) and imaginary part ($\varepsilon_2(\lambda)$) of the dielectric function with respect to wavelength from plot images using WebPlotDigitizer. (a) Original line plot image [28]. The line for $Ge_{15}Sb_{85}$ is to be extracted. (b) Purple color for the $Ge_{15}Sb_{85}$ line plot is specified and the line area is roughly marked with yellow color using mouse input. WebPlotDigitizer then processes the marked area and extracts the specified purple-colored line. (c) Red dots show the (x, y) coordinate values extracted with WebPlotDigitizer.

Later approaches such as ReVision [29] and ChartSense [30] employ computer vision techniques to further automate chart identification process but its performance is in need of further improvements to be utilized reliably for data extraction with low error rates. Davila *et al.* [31] point out that chart data extraction is a challenging process where both text legends and graphics need to be interpreted correctly. Development of a style-independent chart recognition system is another challenge where diverse styles for graphic chart plots exist. Ray Choudhury and Giles [32] approach the data extraction problem from the perspective of a digital library where they intend to construct research article repositories with the availability of metadata and metadata will include data automatically extracted from the charts. While their ideas provide a systematic approach for retrieving datasets from published research articles, its incorporation into current research article repositories (such as a form of adding metadata) will be investigated and likely incorporated in future iterations.

Data extraction from research articles has the potential of collecting high-quality data that already went through validation processes (post-experiment) and has benefited from peer review. However, due to the involvement of time-consuming manual work for chart recognition tools, the advantage of visual graph plot availability can hardly be translated to the recovered original dataset. For this reason, we currently intend to continue using WebPlotDigitizer for data extraction due to its maturity as a reliable software application, even though WebPlotDigitizer requires a time-consuming manual plot area marking process. While current solutions for data extraction from research articles listed above are not sufficient for our purpose, it should be noted that they have the potential to generate reliable, high-quality data once the techniques mature or a large-sized research article repository with metadata becomes available.

**Crowdsourcing**
While the definition of crowdsourcing [33] can vary, we want to narrow it to the collection of material characteristics data for PCMs from contributing researchers and the storage of the submitted data onto our database system. Crowdsourcing is expected to be effective in building a large-scale database, since the influx of experimental data submissions will increase as more number of contributors become available. MetaNet (http://metanet.stanford.edu) [19] is an online database of design algorithms and device layouts for photonic technologies. Currently they provide metasurface device layout and accept submission of data from contributors via e-mail. Similarly to MetaNet, we hope to eventually provide a large-scale database service for PCM data analysis for the research community with our PCM-net system.

To accept experimental data submissions from outside contributors, template-like standardization of the data submission is important to streamline the submission process and manage the data validity and integrity for the database. Esnayra *et al.* [34] list error prevention and correction as important factors for data integrity when collecting crowdsourced bioinformatic genome sequence data. While the data domain is different, it is expected that the PCM-net database system will require similar error prevention during the data submission process and error correction afterwards to maintain the high data quality. However, this quality assurance and integrity checking of crowdsourced data can be a challenge since contributors have low incentive to double-check their data. Complexity of error detection and correction of existing data is expected to increase as the amount and the scale of the submitted data increases.

**Software / Simulation**

The primary function of the WVASE® software is to determine the complex refractive index and thickness of thin-film layers by finding the best match model to experimental data. However, the software contains many other features and functions which we envisage as ideal complements to the PCM database, a few examples are introduced in this section.

1) Simulation of reflectance and transmittance: With the optical constants of interested materials collected in the database, we can simulate intensity-based quantities such as reflectance and transmittance for unpolarized light or user defined polarization states using WVASE®. For optically anisotropic devices which may convert the *p*-polarization to the *s*-polarization and vice versa, the software enables the user to calculate both the polarizing and cross-polarizing intensities. The wavelength range, angles of incidence, and ambient index can be specified as needed. For transparent substrates, the software can account for backside effects associated with the incoherent backside reflections [35] or reverse-side effects where the beam enters from the substrate side. Intensity simulations may be used to predict and engineer device performance, especially when the sample fabrication is challenging or time-consuming.

2) Alloy and temperature files: For compound materials, the database may only have discrete data points with each datum representing a certain composition. Once a set of various composition material files are prepared through measurements, we can create an alloy file in WVASE®. The alloy file is a parameterized material file where the optical constants of a material are described by a function of alloy fraction. Therefore, this material file is used to identify the composition information of unknown samples. For a sample whose alloy fraction is known but exists between two previously measured compositions, the alloy file predicts the optical constants of the material using an interpolation method. WVASE® adopts the energy shift method [36] for the interpolation procedure. This method is useful for compound semiconductors where the critical points associated with electronic transitions and the bandgap energy shift systematically according to alloy fraction. Similarly, we can create temperature files where the control parameter is the measurement temperature, or alloy-temperature files for any materials that show a systematic shift in optical properties according to composition and temperature. Figure 2 shows an example of a temperature file created for C-doped $Ge_2Sb_2Te_5$ (C:GST) at elevated temperatures from 425K (152°C) to 625K (352°C) [26]; added to the database. The complex dielectric constants measured at five different temperatures vary systematically and the generated curves for 550K (277°C), using the energy shift interpolation method, are well placed between the two adjacent data points.

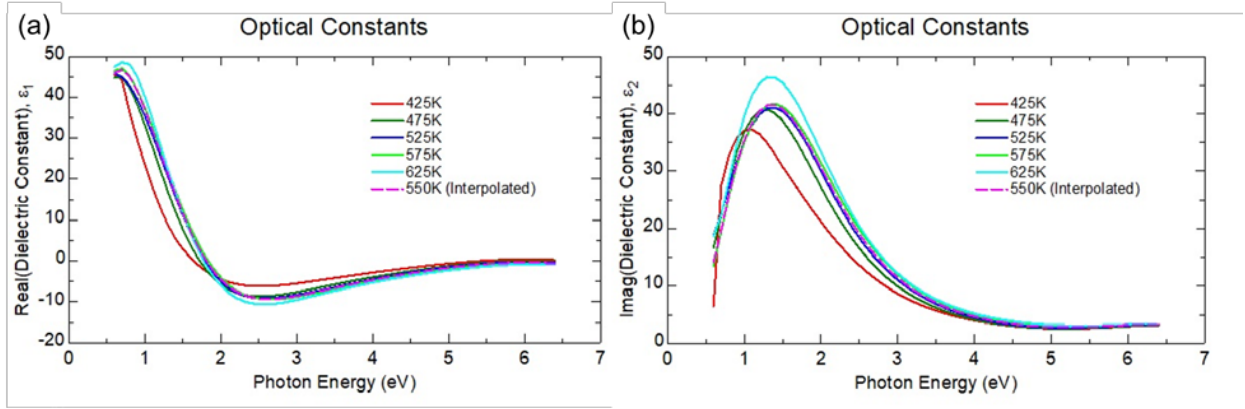

**Figure 2.** Temperature-dependent optical constants of C:GST phase change alloys. The complex dielectric constants, both (a) real ($\varepsilon_1$) and (b) imaginary ($\varepsilon_2$) parts, measured at five different temperatures of 425K (152°C), 475K (202°C), 525K (252°C), 575K (302°C), and 625K (352°C) and the generated curves for 550K(252°C) using the energy shift interpolation method.

3) Effective Medium Approximations: In future work, we will expand the database to create a new set of optical constants by mixing two or more materials. This is of particular interest for partial crystallinities of GeSbTe / otherwise, whereby intermediary states between fully amorphous and fully crystalline states are achievable [7,8]. The average optical property of a mixture may be described by effective medium approximations (EMAs). They work most successfully at the mesoscopic or macroscopic scale where the macroscopic optical properties of a mixture are described as a homogeneous medium while the properties of the constituents may vary inhomogeneously at the microscopic level. To predict the average or effective dielectric function, we can apply EMAs based on the known dielectric function of each constituent and the mixing ratio. WVASE® is capable of mixing two or three materials to predict the effective dielectric constants using the well-known Maxwell-Garnet formula or Bruggeman approximation.

The Maxwell-Garnett EMA formula is given in eq. (1) for isotopically mixed materials with spherical inclusions [37].

$$\frac{\varepsilon - \varepsilon_0}{\varepsilon + 2\varepsilon_0} = \sum i \left( f_i \left( \frac{\varepsilon_i - \varepsilon_0}{\varepsilon_i + 2\varepsilon_0} \right) \right) \quad (1)$$

The effective dielectric constants $\varepsilon$ is calculated by the dielectric constant of a host material $\epsilon_0$ with inclusions of dielectric constants $\varepsilon_i$. The volume fraction of each inclusion is denoted by $f_i$. The main shortcoming of the Maxwell-Garnet formula is that the effective dielectric constants are not identical depending on the choice of the host material. Despite the asymmetry, this theory is well-suited for certain types of composites where the host material takes up most of the volume. The Bruggeman EMA theory overcomes the difficulty of the Maxwell-Garnett theory by treating the two constituents in a symmetrical fashion as given in eq. (2) for spherical inclusions [37].

$$\sum i \left( f_i \left( \frac{\varepsilon_i - \varepsilon}{\varepsilon + 2\varepsilon_0} \right) \right) = 0 \quad (2)$$

The two EMA theories agree well when the host medium is filled with a small volume of inclusions. The Bruggeman EMA is preferred when the concentrations of the mixed materials are comparable or the dielectric constants of them are largely different for its symmetric form.

We can further expand the database by including anisotropic mixing. When the inclusions are non-spherical or the interfaces show preferred directions, the effective dielectric function may reveal optical anisotropy [38]. For anisotropic EMA modeling, we use the depolarization factor to account for electric field screening. For instance, when the light penetrates a mixed material at normal incidence, the depolarization factor is the minimum for vertical interfaces and the maximum for horizontal interfaces. Therefore, anisotropic EMA works for shape induced anisotropy by allowing direction dependent depolarization factor. Some typical examples are stacked bilayer metamaterials [39], nanorods, nanocolumns, spiral films, and slanted needles [40]. A demonstration of anisotropic Bruggeman EMA is shown in Figure 3. We simulated the effective refractive index of a repeated bilayer stack made of $Ge_2Sb_2Te_5$ thin films with two different phases, amorphous and crystalline. The individual layers show isotropic optical constants as shown in Figure 3(a). The amorphous phase is highly absorbing while the crystalline phase is transparent in the long wavelength range. Interestingly, when these two films are stacked repeatedly along the vertical direction and the mixing ratio is 4:1, the effective optical constants are dielectric-like along the z-direction and metal-like in the x-y plane. As long as individual film thicknesses are thinner than the wavelength of interest and the mixing ratio is held fixed, the EMA model predicts indifferent results no matter how thick or thin the films are. This simulation is given to prove the usefulness of EMA modeling in expanding the database and creating novel optical devices.

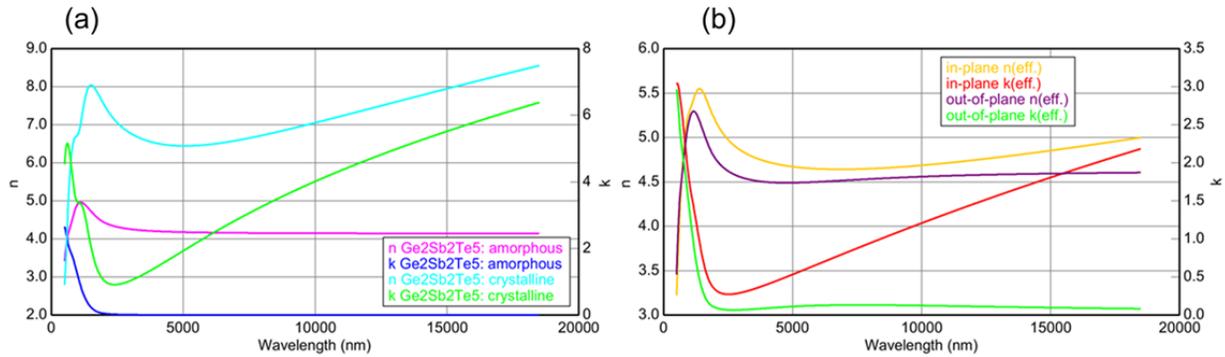

**Figure 3.** (a) Experimentally measured complex refractive index of $Ge_2Sb_2Te_5$ films with two different phases. (b) Anisotropic Bruggeman EMA model simulation for a stacked bilayer of the two materials with the mixing ratio of 4:1 (amorphous to crystalline phase).

## PCM-net Database: Consolidation of the collected data and visualization

To consolidate the data collected from the diverse sources discussed above, a database system is required that organizes collected data with consistency and integrity. Organization of data storage can conceivably start from simple table formats found in spreadsheet applications such as Microsoft Excel. However, spreadsheets present limitations as the scale of data increases and new materials are added since spreadsheets provide limited organization schemes for the data. Graph plot visualizations in spreadsheets need manual configuration of the data area marked for plotting whenever new input data updates are entered. When a new material is added, a separate spreadsheet page will be needed and this will make it difficult to generate graph plot visualizations consistently across multiple spreadsheet pages for different materials. Sharing of the collected data among multiple users can be tricky due to processing of data access privilege and the added metadata such as user accounts.

The retrieval or conditional filtering of data from the spreadsheet pages can become complex beyond the features provided by the spreadsheet applications. While spreadsheets provide filtering schemes for basic tasks, database systems can be a better solution choice since data retrieval conditions can be specified logically using database query languages such as SQL (Structured Query Language). Since the PCM-net system aims to create a large-scale data repository, having a data retrieval scheme that meets user's needs beyond the basic filtering in spreadsheets becomes important. Moreover, data submissions through crowdsourcing will require storing additional information such as contributor's profile information and additional metadata. This requirement expedites the adoption of database systems over the use of spreadsheets in order to store the data for PCM-net.

**Database design considerations**
Our PCM-net database will be designed to operate as an independent, separate layer replaceable with other data solutions. While MySQL relational database is selected as the database solution for our first-pass prototyping purpose, we think that other databases can be plugged-in instead of MySQL in case MySQL should show any limitations. One disadvantage of relational database systems is that it must have a fixed data schema design from the outset. In relational databases, it is usually difficult to change the data schema once its design is fixed. It might lead to redesign of the entire database in case the current data schema fails to incorporate new input data types. This inflexibility is being addressed [41] in other non-relational databases such as NoSQL databases, where JavaScript Object Notation (JSON) format provides an alternative for document-oriented data storage without predetermined database schema. We plan to adopt NoSQL database if MySQL relational database meets short of our expectations in the PCM-net system.

**Database application development**
The relational database design starts from a simple table structure where combinations of wavelength, refractive index, extinction coefficient, material state (amorphous, partial crystallization, full crystallization), and the phase change material. Using the normalization technique of typical relational database design, the Material table is separated out to accommodate new additional phase change materials without duplicates and model one-to-many relation between the material and the rest of the data. (i.e. one material can have many combinations of wavelength, refractive index, extinction coefficient, and material state). Figure 4 shows the entity-relationship (ER) diagram with the attribute definitions for Material and Data table. Refractive index and extinction coefficients are represented respectively as *n* and *k* attributes. amorphousToCrystalline attribute denotes the material state, where 0 is for amorphous state, 1 is for crystalline, and any value between 0 and 1 is for the intermediate state as partial crystallization. Figure 4(d) shows an example of table data entries for $Ge_2Sb_2Te_5$. It can be seen that the Data table stores all the particular wavelength, *n*, *k*, and amorphousToCrystalline data pairs, with the materialId points to the PCM referenced back in the Material table.

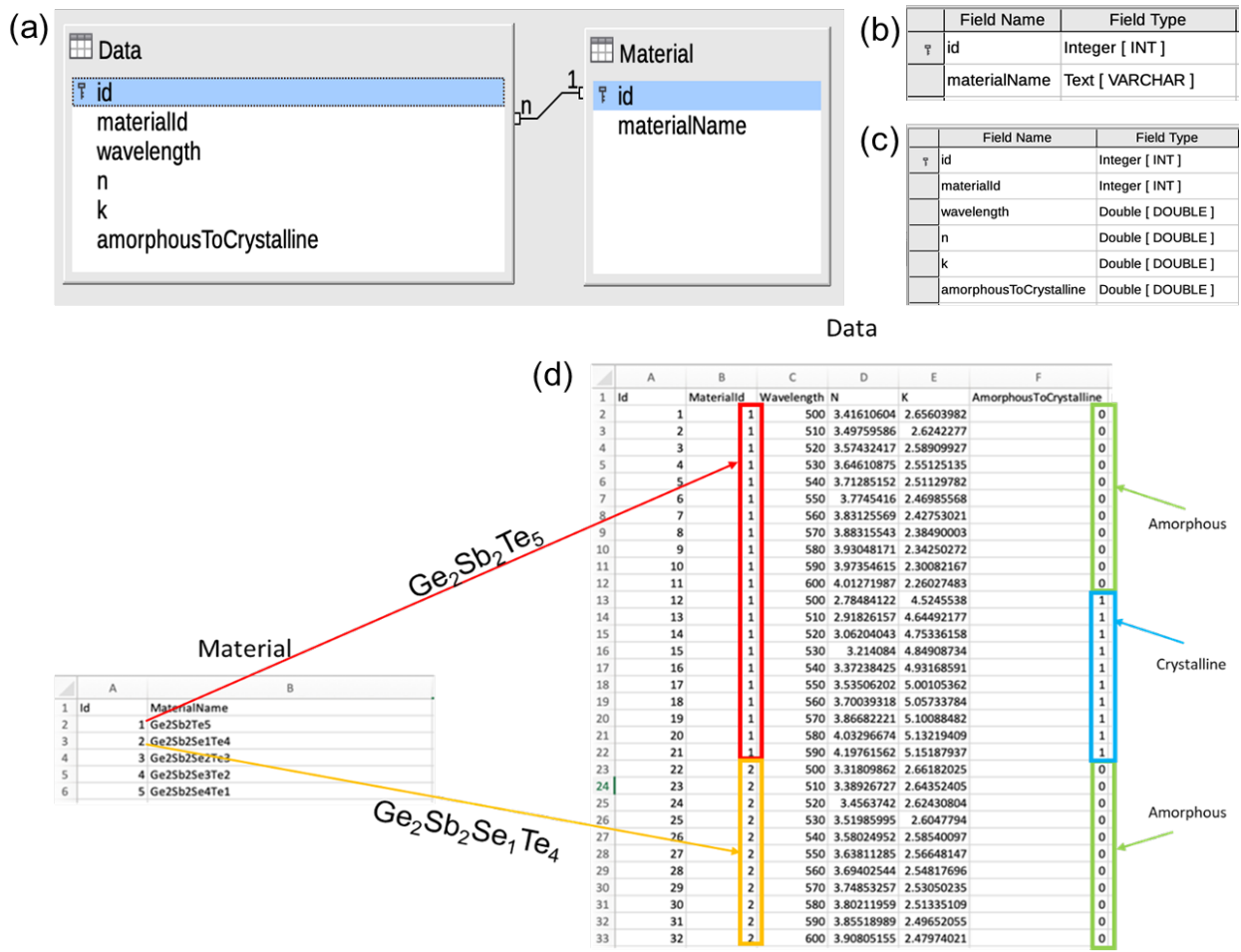

**Figure 4.** (a) The entity-relationship (ER) diagram for Data and Material tables. They are linked with 1:*n* relationship using unique id attribute. (b), (c) Data type definitions of the attributes in Material and Data tables. (d) An example of table data entries for $Ge_2Sb_2Te_5$ and $Ge_2Sb_2Se_1Te_4$. $Ge_2Sb_2Te_5$ is assigned with id 1 in Material table and associated with multiple rows in the Data table via materialId. Material states are denoted as 0 for amorphous and 1 for crystalline. Intermediate states can be denoted with numbers between 0 and 1.

While this is a simplistic relational database design bounded by the requirement of predefined data schema, note that the table design has extensibility for new data types. For example, attributes such as operating temperature can be added to the Material table in future work. To implement submissions from crowdsourcing, Data table may need to have additional attributes such as UserId and SubmissionDate. DataGenerationMethod attribute can also be added that specifies whether the data are obtained from direct experiments, extracted from graphical plots (images) in published articles, or generated from simulations (theory). For the time being, we do not consider incorporations of User table and its attributes for multi-user crowdsourcing to keep the database design simple and manageable.

With the initial prototype data being available in the database, users can retrieve the desired dataset with SQL statements that describe the data retrieval conditions. Currently, our prototype database has a total of 29,664 entries in Data table, with 24 types of phase change materials ($Ge_2Sb_2Te_5$, C-doped $Ge_2Sb_2Te_5$, $Ge_{15}Sb_{85}$, $Sb_2S_3$, $Ge_2Sb_2Se_xTe_{5-x}$ (x = 0 to 5), $Ge_2Sb_2Se_4Te_1$, $Ge_2Sb_2Se_5$, and so on) listed in Material table. All the refractive index, extraction coefficient, and amorphous/crystalline state data are extracted from graph plot images in peer-reviewed research articles (also referenced on the database, http://nekocloud.com/pnet/) using WebPlotDigitizer. Once the database is ready, users can retrieve the desired dataset with the filtering commands (such as SQL). Several examples are listed below and the interactive line plots in the next section are based on these examples:
1. Characteristics of a selected phase change material:
    - Retrieve (wavelength, *n*, *k*, AmorphousToCrystalline) values for a selected phase change material
2. Comparison across phase change materials:
    - Find (wavelength, *n*, *k*) values grouped by its amorphous/crystalline state across all the phase change materials
3. Further data processing:
    - Find the differences between *n* and *k* values in amorphous or crystalline state for a selected material

**Data visualization**

Data visualization, especially the relationship between refractive index and extinction coefficient with wavelength, is a useful tool for understanding the characteristics and performance of PCMs. For our work, Plotly Dash [42] is chosen as the graphical visualization tool. Plotly Dash is a graphical plotting tool written in Python and generates Web-based data visualization plots. Interactivity in graphical plots, and the availability of adding user interface components (such as a drop-down box for selecting phase-change material), are useful features in which Plotly Dash shows its strength. One downside of Plotly Dash is that it does not provide any integrated database access features and the component for data retrieval needs to be manually implemented. On the other hand, while other database front-end applications such as Microsoft Access or LibreOffice Base provide integrated database access and user interfaces for managing data stored in the database, their graph plotting feature is not sufficient to meet our needs. DBFace (http://dbface.com) was considered once since it provides front-end user interfaces and multi-user design that will be useful for future data crowdsourcing implementation from the start. But it was dropped for the same reasons as the front-end applications since its graph plotting features are not sufficient for our needs.

Figure 5 shows the graphical line plots for the dataset retrieved in the previous section. Note that the plot provides user control on material selections or amorphous / crystalline states by providing graphical user interface (GUI) components and visual plotting of the *n* or *k* values vs. wavelength. We plan to add similar plot pages that further meet users' needs and build the collections as an application. Currently the prototype data plot pages can be accessed from the URL (http://nekocloud.com/pnet). Users can choose options such as phase state, material, and additional preferences via user interface components. The line plots show the differences between amorphous and crystalline phases of *n* and *k* values (*Δn* and *Δk*) respectively. The overlapped purple region in Figure 6 shows high *n* and negligible *k*. The spectral characteristics of the PCM database can be utilized in order to determine optimal operating wavebands (windows) for each PCM, which is useful for spectroscopic applications where low loss material selection is critical for efficient tunable optical filters. In addition to visualization, the charts have features such as exporting the index data (for external usage in simulation packages / otherwise) and downloading the plots in different image file formats.

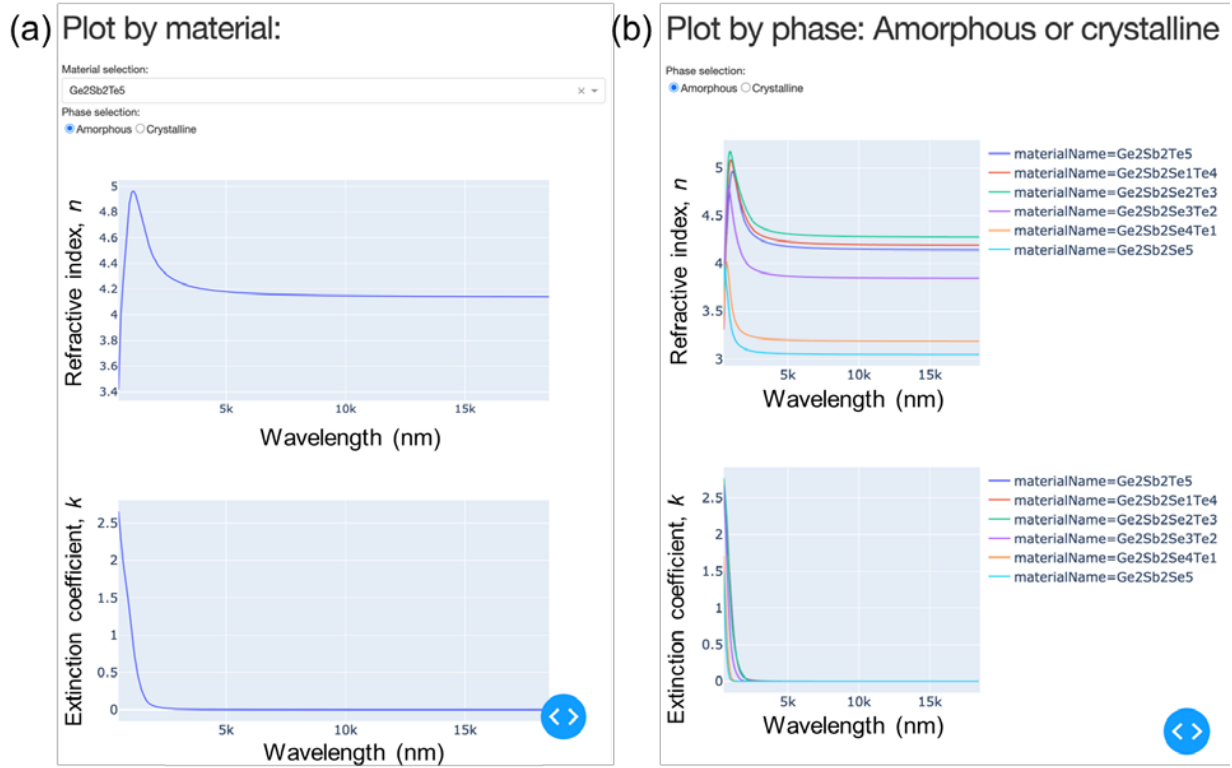

**Figure 5**. PCM-net data visualization examples generated with Plotly Dash. (a) plot by material and (b) plot by crystal phase of PCM.

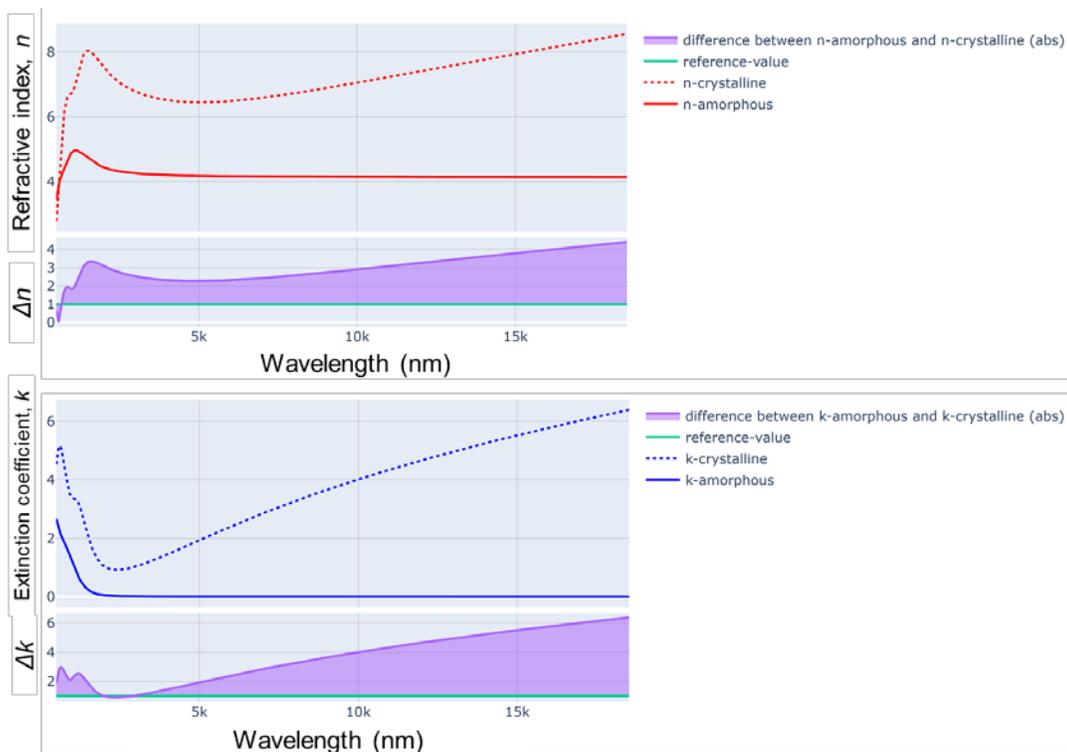

**Figure 6.** Data visualization examples in PCM-net generated with Plotly Dash of optical constants ($n$ and $k$) difference between amorphous and crystalline phases.

**Future implementation plans**
Our database is in its infancy, however its utility and appeal to the wider community should be apparent. In order to implement crowdsourced data from contributors, the database schema of the PCM-net database system has to be extended to add contributor's profile information (such as sign-in account, encrypted password, and contact information) and metadata (data generation method, experiment name, date, and other information). In our current MySQL-based relational database schema, metadata attributes will be added to the Material table. A separate User table will be needed to store contributor's sign-in account, profile information, and details on data contributions. When the data schema is prepared, the PCM-net database system can evolve into a web-based, multi-user database application. Implementation of user authentication features, data view, and data submission pages will be needed. As an interim step before full implementations, a contribution data submission template in spreadsheet format, such as popular Microsoft Excel, is to be trialed.

Spreadsheet-based data submission templates are expected to have the following advantages: First, spreadsheet applications have user-friendly interfaces and naturally become the application of choice for researchers to store any number data. Second, the table-based format, while it does not provide top-level data organization schema, can be easily parsed with scripting languages such as Python and converted to MySQL-based relational database table entries. Third, basic input data validation and unit checking can be implemented in the spreadsheet template file by embedding scripting code in the spreadsheet applications. This will make first-pass input data validation possible before the contributor actually submits the data to our system and is expected to simplify second-pass data quality check steps on the submitted data. Once the spreadsheet-based submission template design is complete, a new user interface for the data submission can be designed based on the template. Finally, the graphical plot (visualization) scripting code for Plotly Dash pages will need a full-featured database access component layer for the backend PCM-net database stored in MySQL. Note that Plotly Dash is a plotting tool and does not include any database-related solutions in itself. This gap has to be filled by implementation of the database access component and use it for data retrieval from the PCM-net database.

# PCM-net benefits for
# tunable nanophotonic / metasurface device modelling

The refractive index database of chalcogenide PCMs from PCM-net is beneficial for tunable nanophotonic device modelling, in which both knowledge of waveband-appropriate PCM to use and its subsequent accurate index values are required. Tunable nanophotonic / metasurface devices, where the properties of the meta-atoms can be tuned to modulate the response of the optics, enable multifunctional operation capabilities (variable zoom, tunable spectral filtering, and / or field-of-view (FOV) steering) and further enable compact and rugged optics with reconfigurability [43]. Figure 7 shows an overview of potential applications of PCM-based tunable nanophotonic / metasurface imaging optics and hence fields which would benefit from our collated PCM index database. For example, tunable infrared imaging optics provides a core technology for health monitoring, aerial vehicles, air pollution in emission, or absorption patterns for air-quality monitoring [44,45]. Tunable PCM-based metasurface optics is also useful in photovoltaics / thermophotovoltaics due to its autofocus feature with zoom lens capabilities. PCM metasurface optics comprising arrays of sub-wavelength antennas offer a promising solution for high speed communications and provide significant SWaP-C (Size, Weight, Power, and Cost) advantages for miniaturized spectroscopy in biomedical and space applications.

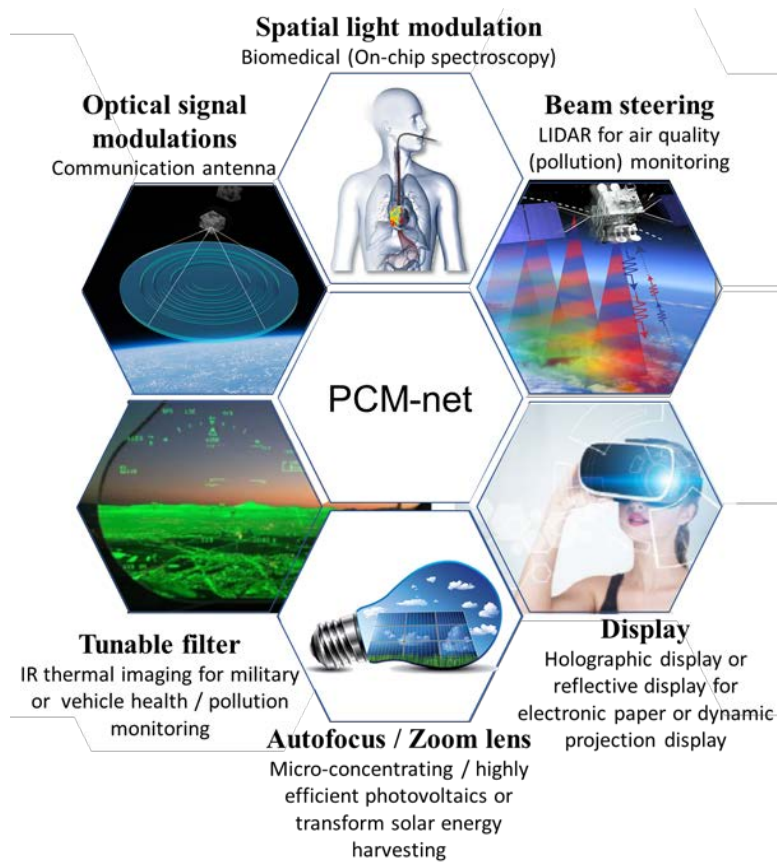

**Figure 7.** Various applications of PCM-based tunable nanophotonic devices. The optical properties collated and available from PCM-net are beneficial for a range of application-driven device modelling.

**Application example at NASA: PCM tunable filters**
Tunable infrared (near to mid-wave) filters provide a core technology for health monitoring of aerial vehicles, probing molecular vibrations in chemical species to detect radiant thermal signatures [44,45]. The potential SWaP-C advantages of PCM-based tunable filters over traditional approaches, such as bulky filter wheels or expensive acousto-optical tunable filters, along with filtering capability of ultrafast tuning across wide wavebands, makes PCM-based optical filter technologies highly attractive for current and future applications at NASA.

PCMs are largely transparent across various spectral wavebands and exhibit significantly large, reversible refractive index modulation upon crystallization, making them ideal candidates for use in optical spectrally robust filter designs [14-15]. PCM-net collates and provides dataset plots of this information and Figure 8(a) shows the waveband-appropriate PCMs based on a minimal extinction coefficient consideration and the large figure of merit, from Figure 6. A large figure of merit ($\Delta n/\Delta k$) and low extinction coefficient ($\Delta k \approx 0$) between amorphous and crystalline phases are critical for identifying the most appropriate PCM for the waveband, and application, of interest. One or multiple PCMs can be selected for tunable filter modelling depending on the targeted wavelength (or waveband). PCM-net is leveraged in the design and simulation of several tunable devices utilized in multiple NASA program scenarios, which all have varying, but stringent, performance requirements. At present, these missions, which are utilizing PCM-net, are: DIAL (Differential Absorption Lidar) science missions, the SAGE (Stratospheric Aerosol and Gas Experiment) mission, and SCIFLI (Scientifically Calibrated In-Flight Imagery) program; highlighted in Figure 8(b) and discussed in further detail below:

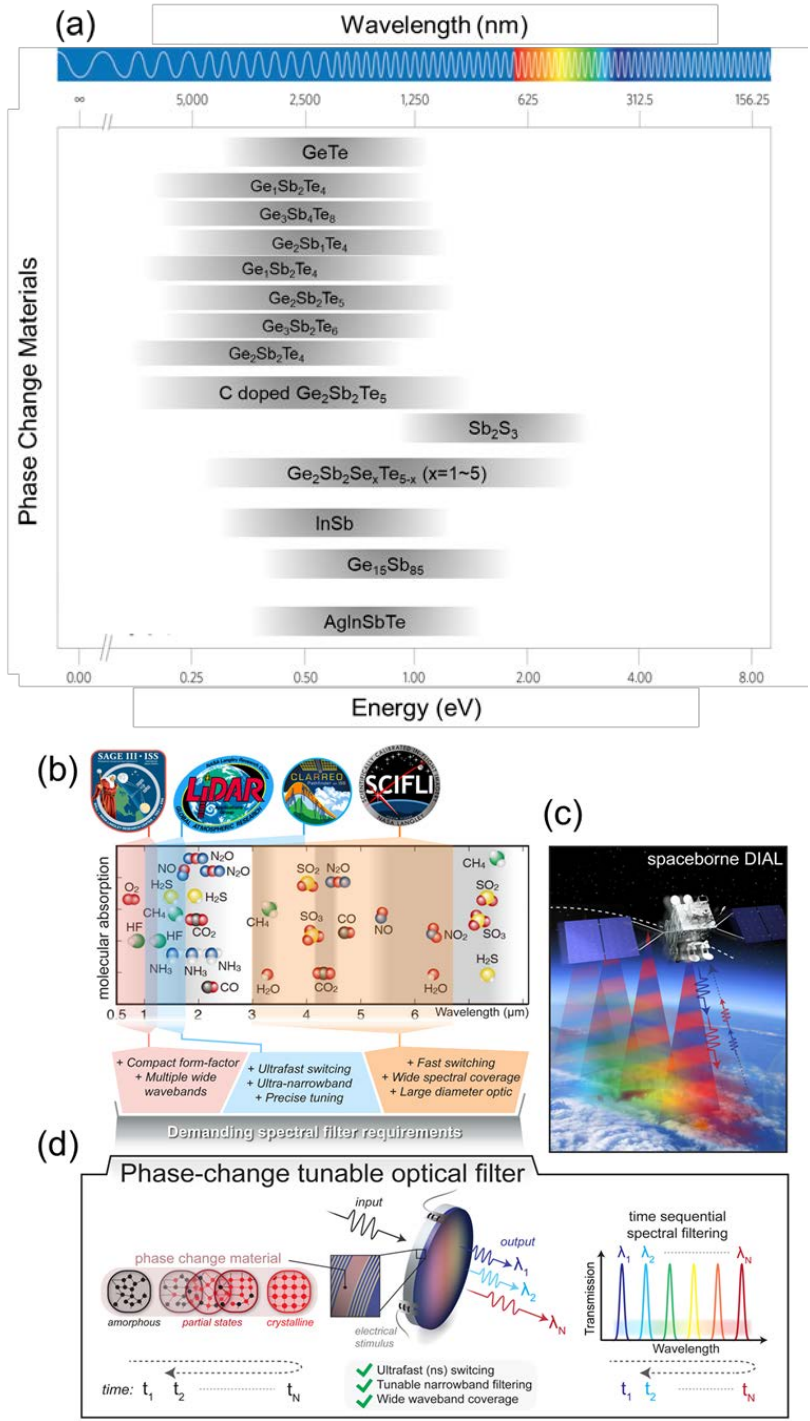

**Figure 8.** (a) Plot of suitable wavebands for different PCMs based on extinction coefficient consideration and figure of merit. (b) Phase-change tunable optical filters for NASA applications based on targeted (operational) wavelength (the molecular absorption image has been adapted from Hamamatsu, *https://www.hamamatsu.com/us/en/support/inquiry/index.html*, and edited by authors). (c) SmallSat-based Differential Absorption LIDAR (DIAL) operating schematic (d) The technology of the Phase Change Material (PCM)-based tunable spectral filtering.

Differential absorption LIDAR (DIAL) is a strategically important measurement technique within NASA, mapping range-resolved concentrations of trace/greenhouse gases such as, $H_2O$ (~890 nm wavelength) and $CH_4$ (1.6 µm wavelength) in the atmosphere which are important to many processes that underpin weather and climate systems [44]. Spaceborne DIAL would provide for the direct and unbiased profiles of gases throughout the atmosphere with high vertical resolution and global coverage beyond the existing airborne systems [Figure 8(c)]. The benefit of developing a PCM-based tunable filter is that only one filter and one detector are required regardless of the number of wavelengths needed to transmit for sampling the gases at various spectral locations. This is in direct contrast to the established approach of using a combination of bulky optical components (beamsplitters, filters, etalons etc.) which as a system has a large footprint, power consumption and has demanding alignment tolerances. A single element tunable filter (PCM-based) allows SmallSat-based spaceborne DIAL to decrease SWaP-C, risk, and reduce systematic bias that could result from non-linear time dependent degradation of the different detectors. PCM-net offers the ability to quickly, easily and accurately identify the optimal PCM for DIAL sub-system filter design.

From Figure 8(a), $Sb_2S_3$, and $Ge_2Sb_2Se_xTe_{5-x}$ (GSST) can be selected as a material candidate for the tunable filter design for DIAL missions. The stoichiometry of the chalcogenide PCM materials is chosen for the targeted spectral ranges: DIAL for 800 nm – 1.6 µm to discern specific chemical species in the atmosphere. In detail, $Sb_2S_3$ is a low-loss phase change material that can switch from a low-index ($n = 3$), amorphous state to a high-index ($n = 4.5$), crystalline state with low loss ($k < 0.005$) over a wavelength range of 600 nm – 2 µm or more [46]. $Ge_2Sb_2Se_1Te_4$ (GSST), has recently been reported broadband transparency (1 – 18.5µm) and large optical contrast ($\Delta n = 2.0$) and proposed as an entirely new range of infrared and thermal photonic devices [12]. $Ge_2Sb_2Te_5$ (GST) has a refractive index of ~3 and ~5.5 in its amorphous and crystalline states ($\Delta n > 2$), respectively across the 2 – 10µm spectral range with negligible extinction coefficient in crystalline phase ($\Delta k \approx 0$) but high loss in its crystalline phase when compared to $Sb_2S_3$ and GSST, which has utility in transmissive mid-wave infrared applications only, not for DIAL applications [47,48]. $Sb_2S_3$ and GSST can be switched thermally, optically with visible light, or electrically via joule heating at speeds from MHz to GHz. Through the integration of these PCMs as the active material in thin-film optical filter designs, all-solid-state spectrally tunable optical filters are achievable. Each PCM state (i.e. amorphous, partial crystallinities, or fully-crystalline) has an associated varying refractive index [49-51], and through PCM-net influenced design, can be utilised as the tuning medium which controls the passband center wavelength upon energy stimulus (Figure 8d).

## Conclusion

The growing demand for increasingly functional photonic devices has seen phase change materials (PCMs) cement themselves as attractive optically active materials for a diverse range of applications. Owing to their large optical contrast arising from a non-volatile phase transition, PCMs can be integrated as the tunable medium in different optical device designs to provide ultrafast modulation (wavelength, amplitude, phase) across multiple wavebands. Designing and simulating multifunctional devices based on PCMs, such as GeSbTe, is challenging; optical performance metrics must be optimized at multiple optical phase states and accurate representation of the optical constants of various PCMs is critical. We have introduced PCM-net (http://nekocloud.com/pnet/): a prototype online database of chalcogenide PCMs' optical characteristics as an accessible and indexed repository for the research community. Refractive index ($n$) and extinction coefficients ($k$) of PCMs between amorphous and crystalline states are collected, and indexed, from various sources. Using experimental data directly extracted from academic research articles, and submitted by research contributors, we have thus far made available over twenty PCM datasets to the database and plan to grow this database with community collaboration over forthcoming years. The data collecting approaches are presently sufficient for the users' needs, however an advanced data collection process is required as the database grows and user base expands, and we envisage the data computationally generated by commercial

WVASE® software, the ellipsometry analysis program. We have further shown an application example of PCM-net: PCM selection and optimization for tunable optical filters for current and future missions at NASA, which include specific scenarios of the DIAL system and SCIFLI program. We anticipate that the coordinated research efforts enabled by PCM-net will, (1) expedite sufficient resources sharing of the PCM data, and (2) promote repositories with an aim to select the most appropriate PCM for targeted real-world applications including tunable metasurface optics, integrated photonic waveguides, and ultrafast optical modulators.

## Funding

NASA Langley Research Center Internal Research & Development Program (IRAD); Space Act Agreement (SAA) between NASA and University of Cambridge, cooperative research on 'Database development for the optical resonance design'; Engineering and Physical Sciences Research Council (EP/R003599/1); Wellcome Trust (Interdisciplinary Fellowship).

## Acknowledgement

The authors acknowledge the PCM-based tunable metasurface collaboration team, Prof. Juejun Hu (MIT), Dr. Tian Gu (MIT), Dr. Matthew Julian (Booz Allen Hamilton), Dr. Amin Nehrir (NASA LaRC), Dr. John Smith (NASA LaRC), Dr. Rory Barton-Grimley (NASA LaRC), Mr. Carey Scott (NASA LaRC), Mr. David MacDonnell (NASA LaRC), Mr. Scott Bartram (NASA LaRC), and Mr. Stephen Borg (NASA LaRC). H.J.K. would like to acknowledge the support of Mr. Thomas Jones at NASA LaRC. C.W. would like to acknowledge the support of Prof. Sarah Bohndiek.